\documentclass[a4paper,12pt]{article}
\usepackage{array}
\usepackage{amsmath}
\usepackage{graphicx,subfigure,color}
\usepackage{epsfig}
\usepackage{fullpage}
\usepackage[normalem]{ulem}
\usepackage{amsfonts}
\usepackage{multirow,multicol,lscape}
\usepackage{amssymb,bm}
\usepackage{amsmath}
\usepackage{natbib}%
\usepackage{indentfirst}
\usepackage[all]{xy}
\usepackage{enumitem}
\usepackage{placeins}

\numberwithin{equation}{section}

\usepackage{natbib}

\newcommand{\bftheta}{\hbox{\boldmath$\theta$}}
\newcommand{\bfy}{\hbox{\boldmath$y$}}

\begin{document}

\title{Bayesian Transformed GARMA Models}

\author{
  Breno S. Andrade$^a$, 
  Marinho G. Andrade$^b$, 
  Ricardo S. Ehlers$^b$\\
  {\small $^{a}$Programa de P\'os-Gradua\c{c}\~ao Interinstitucional UFSCar/ICMC-USP}\\
  {\small $^{b}$Department of Applied Mathematics and Statistics,
    University of S\~ao Paulo}\\
}

\date{}

\maketitle

\begin{abstract}

Transformed Generalized Autoregressive Moving Average (TGARMA)
models were recently proposed to deal with non-additivity, non-normality and
heteroscedasticity in real time series data. In this paper,
a Bayesian approach is proposed for TGARMA models, thus extending the
original model. 
We conducted a simulation study to
investigate the performance of Bayesian estimation and Bayesian model
selection criteria. In addition, a real dataset was analysed using the
proposed approach. 


\vskip .5cm

\noindent{\bf Keywords}: {\it Generalized ARMA model, Bayesian
  inference, Transformations, gamma distribution, inverse Gaussian
  distribution} 

\end{abstract}

\section{Introduction}

Generalized Autoregressive Moving Average (GARMA) models extend the
classical ARMA time series models and have been
around since the seminal work of \cite{Benjamin}. GARMA models are designed
to accommodate time series data (either discrete or continuous)
associated with distributions in the exponential family. However, to
ensure that the usual assumptions for GARMA models hold a
transformation of the original data may be necessary. This is the idea
behind the so called Transformed
Generalized Autoregressive Moving Average (TGARMA) models proposed
recently by
\cite{tgarmas} which provide a great deal of flexibility in modeling
time series data with possible
non-additivity, non-normality and heteroscedasticity. The authors
applied maximum likelihood and bootstrap methods for parameter
estimation and prediction.
The main motivation for the present work is to provide a fully
Bayesian approach to estimate parameters, compare models and make
predictions in TGARMA models. 

In the recent literature, transformations have been shown as a good
alternative to reduce certain anomalies 
in the data. For example, \cite{hamasaki} described a Box and Cox
power-transformation to confined and censored nonnormal responses in
regression, \cite{silva} proposed the use of Box-Cox transformations
and regression models to deal with fecal egg count
data and \cite{gillard} presented a study using a Box-Cox family of
transformations and highlight problems with asymmetry in
the transformed data. \cite{castillo} commented about many fields
where the Box-Cox transformation can be used, and also proposed a
method to improve the forecasting models. \cite{ahmad} combined
Box-Cox transformation and bootstrapping ideas in one single algorithm
where the transformation was used to
ensure the data is normally distributed while bootstraping allowed to deal with
small and limited sample size data.

\cite{tgarmas} proposed the estimation of the transformation parameter
$\lambda$ via profile likelihood (PL). \cite{zhu2} presented a
procedure to dimensionality selection maximizing a profile likelihood
function.  \cite {huang} proposed an efficient equation for estimating
the index parameter and unknown link function using adaptive
profile-empirical-likelihood inferences. Finally, \cite{cole} provide a primer
on maximum likelihood, profile likelihood and penalized likelihood
which have proven useful in epidemiologic research.

In the Bayesian approach, computational methods based on Markov chain
Monte Carlo (MCMC) can be utilized to address the complexity of the profile
likelihood. Thus, our main contribution concerns inference under the
Bayesian framework providing efficient MCMC procedures to evaluate the joint
posterior distribution of model parameters. Recently, \cite{brenobayes}
presented a Bayesian approach for GARMA models, indicating advantages
of using Bayesian methods. This paper builds upon previous work
and extends the fully Bayesian approach to transformed GARMA models by
assingning a prior distribution to the transformation parameter which
is then embedded in the estimation process. Prior constraints on
$\lambda$ are easily incorporated and these are guaranteed to be valid
in the posterior distribution. We also use Bayesian information
criteria to compare competing models. Finally, properties of MCMC were
used to improve predictions and construct prediction intervals.

The rest of the paper is organized as follows. In Section
\ref{sec:garma} we present the main concepts associated with TGARMA
models. Section \ref{sec:bayes} defines the Bayesian approach for this
new class of models. Section \ref{simulations} contains the simulation
study and Section \ref{real} presents the real data analysis on
Swedish fertility rates. Finally, Section \ref{conclusions} gives
concluding remarks.

\section{Transformed Generalized Autoregressive Moving Average (TGARMA) Model}
\label{sec:garma}

\cite{BoxCox} commented that many important results in statistical
analysis follow from the assumption that the population being sampled
or investigated is normally distributed with a common variance and
additive error structure. For this reason, these authors presented a
transformation called Box-Cox power transformation that has generated
a great deal of interests, both in theoretical work and in practical
applications. This family has been modified by \cite{Cox2} to take
account of the discontinuity at  $\lambda=0$, such that, 
\begin{displaymath}
y_t^{(\lambda)} = \left\{ \begin{array}{ll}
 \frac{(Y_{t}^{\lambda} -1)}{\lambda};  \lambda \neq 0 \\
 \log(Y_{t});  \lambda = 0
\end{array} \right.
\end{displaymath}

\cite{Sakia}, \cite{Manly} and \cite{DraperCox} discuss others
transformation which have the same goal, namely to reduce anomalies in the
data. In most applications, the literature recommends the use of Box-Cox power
transformation as a general transformation. In the next section we present
the TGARMA approach to time series data using this Box-Cox power
transformation in conection with the exponential family of distributions.

\subsection{Model definition}

The TGARMA model specifies the conditional distribution of each
transformed observation $y_t^{(\lambda)}$, for $t=1,\ldots,n$ given
the previous information set, defined as ${F}_{t-1}^{(\lambda)} =
(y_{1}^{(\lambda)},\ldots,y_{t-1}^{(\lambda)},\mu_{1},\ldots,\mu_{t-1})$. This
conditional density belongs to exponential family and is given by,

\begin{equation}\label{defgarma}
f(y_t^{(\lambda)}|{F}_{t-1}^{(\lambda)}) =
\exp\left(\frac{y_t^{(\lambda)}\alpha_t - b(\alpha_t)}{\varphi} +
d(y_t^{(\lambda)},\varphi)\right), 
\end{equation}
where $\alpha_t$ and $\varphi$ are the canonical and scale parameters,
respectively. Moreover $b(\cdot)$ and $d(\cdot)$ are specific
functions that define the particular exponential family. The
conditional mean and conditional variance of $y_t^{(\lambda)}$ given
${F}_{t-1}^{(\lambda)}$ are represented as,
\begin{eqnarray*}\label{mv}
\mu_t = b'(\alpha_t) &=& E(y_t^{(\lambda)}|{F}_{t-1}^{(\lambda)}) \\\nonumber    
 Var(y_t^{(\lambda)}|{F}_{t-1}^{(\lambda)}) &=& \varphi{b''}(\alpha_t),
 ~t=1,\ldots,n.
\end{eqnarray*}

Following the Generalized Linear Models (GLM) approach the conditional
mean
$\mu_t$ is related to the linear predictor $\eta_t$ by a twice differentiable
one-to-one monotonic function $g$, called $link$ $function$. In
general, a set of covariates {\bf x} can be included into the linear predictor.
However, our main interest here is to take into account time series
features of the data and this is accomplished by adding additional
components allowing for autoregressive moving average terms. In such a case our
model will have the following form,
\begin{equation}\label{predii33}
g(\mu_t) = \eta_t = x'_t\beta +
\sum_{j=1}^p\phi_j\{g(y_{t-j}^{(\lambda)}) - x'_{t-j}\beta\} +
\sum_{j=1}^q\theta_j\{g(y_{t-j}^{(\lambda)}) - \eta_{t-j}\}. 
\end{equation}
where the model orders $p$ and $q$ are identified using Baysian information criteria. 
The TGARMA($p,q$) model in its most general form is defined by
equations (\ref{defgarma}) and 
(\ref{predii33}). We note however that it may be necessary to
replace $y_t^{(\lambda)}$ with some $y_t^{(\lambda^{new})}$ in
(\ref{predii33}) to avoid the possible non-existence of
$g(y_t^{(\lambda)})$ for certain values of $y_t^{(\lambda)}$.

In this paper we will consider Box-Cox transformations in two
important continuos GARMA models: gamma and 
inverse Gaussian. Both the simulation study and real data analysis were
performed for each of these distributions. Also, we will not include
any covariate in the linear predictor. Then, the conditional densities
in terms of the mean $\mu_t$ are given by,
\begin{equation*}
f(y_t^{(\lambda)}|{F}^{(\lambda)}_{t-1}) = \frac{1}{\Gamma(\nu)}\left(
\frac{\nu}{\mu_t}  \right)^{\nu}{y^{(\lambda)}_t}^{(\nu-1)}\exp\left(-
\frac{{y^{(\lambda)}_{t}}{\nu}}{\mu_t} \right), ~y^{(\lambda)}_t>0, ~\nu>0,
\end{equation*}
for the Gamma TGARMA model and,
\begin{equation*}
f(y^{(\lambda)}_t|{F}^{(\lambda)}_{t-1}) = \exp\left\{
\frac{1}{\sigma^2}\left[  -\frac{2y^{(\lambda)}_t}{\mu_t^2} +
  \frac{1}{\mu_t} \right]
-\frac{1}{2}\log(2\pi\sigma^{2}{y^{(\lambda)}_{t}}^3)
-\frac{1}{2\sigma^{2}y^{(\lambda)}_t} \right\},
\end{equation*}
for the Inverse Gaussian TGARMA model. In both cases, we use a
logarithmic link function instead of the canonical ones so that the
mean is related to the linear predictor as,
\begin{equation}\label{eq4yt}
\log(\mu_t) = \beta_0 + \sum_{j=1}^{p}\phi_j\left\{
\log(y^{(\lambda)}_{t-j}) \right\} +
\sum_{j=1}^{q}\theta_j(\log(y^{(\lambda)}_{t-j}) - \log(\mu_{t-j})). 
\end{equation}
For estimation purposes we actually used
$y_{t-j}^{(\lambda^{new})} = \max(y_{t-j}^{(\lambda)},c)$,
$0 < c < 1$ in (\ref{eq4yt}), but the superscript {\it new} is dropped
for simplicity.

\section{Bayesian Approach to TGARMA Models}\label{sec:bayes}

To conduct our Baysian analysis we need to specify prior probability
distributions to all unknown quantities.
\cite{BoxCox} showed that the transformation parameter has good statistical
properties when it belongs to the [-1,1] interval. We then assign a
prior distribution for $\lambda$ restricted to this interval. In the
lack of further prior information we specified $\lambda\sim U(-1,1)$,
i.e. a uniform prior distribution. Also, the parameter
$\beta_0$, $\Phi=(\phi_1,\ldots,\phi_p)$ and
$\Theta=(\theta_1,\ldots,\theta_q)$ are assumed a priori independent
and assigned multivariate normal distributions, i.e.
$\beta_0\sim N(\mu_0,\sigma_0^2)$,
$\Phi\sim N(\boldsymbol{\mu_1},\sigma_1^{2}\boldsymbol{I_1})$ and
$\Theta\sim N(\boldsymbol{\mu_2},\sigma_2^{2}\boldsymbol{I_2})$,
where $\boldsymbol{\mu_1}$ and $\boldsymbol{\mu_2}$ are
vectors with lengths $p$ and $q$ respectively;
$\sigma_0^{2}$, $\sigma_1^{2}$ and $\sigma_2^{2}$ represent the prior
variances with $\boldsymbol{I}_1$ and $\boldsymbol{I}_2$ denoting $p$
and $q$-dimensional identity matrices, respectively. So, we also do
not impose prior correlations among the GARMA coefficients as is
common practice. For all models we specified relatively flat priors by
setting relatuvely large values to the variances.

Finally, for the positive parameters $\nu$ and $\sigma^{2}$ in the
gamma and inverse Gaussian cases we propose
lognormal prior distributions, i.e. $\nu\sim LN({\mu_3},\sigma_3^{2})$
and $\sigma^{2}\sim LN({\mu_4},\sigma_4^{2})$.
The hyper-parameters $\mu_3$, $\mu_4$, $\sigma_3^{2}$ and $\sigma_3^{2}$
can also be specified to represent weak prior information.

For a time series of size $n$ with observations
$\bfy=(y_1,\dots,y_n)$, the partial likelihood function is constructed
using (\ref{defgarma}), 
\begin{eqnarray*} \label{q1}
L(\beta_0,\Phi,\Theta,{u}|\bfy) 
&\propto& \prod_{t=r+1}^{n}f(y_t^{(\lambda)}|F_{t-1})  \nonumber\\
&\propto&
\prod_{t=r+1}^{n}\exp\left(\frac{y_{t}^{(\lambda)}\alpha_{t} -
  b(\alpha_{t})}{\varphi} + d(y_{t},\varphi)\right),
\end{eqnarray*}
where the parameter $u$ on the left hand side represents $\nu$ for
the gamma distribution and $\sigma^{2}$ for the Inverse Gaussian one.
Also $\alpha_t = \log(\mu_t)$, which represents the link
function given by (\ref{eq4yt}).

The posterior density is obtained combining the likelihood
function with the prior densities via Bayes theorem. After observing
the time series $\bfy$ the posterior density is then given by,
\begin{equation}\label{jjj1}
\pi(\beta_0,{\Phi},{\Theta},{u},{\lambda}|\bfy) \propto
L(\beta_0,{\Phi},{\Theta},{u}|\bfy)~\pi_{0}(\beta_0,{\Phi},{\Theta},{u},{\lambda}),
\end{equation}
where $\pi_0(\cdot)$ denotes a joint prior distribution.
However, the joint posterior density of parameters can not
be obtained in closed form, therefore Markov chain Monte Carlo (MCMC)
sampling strategies will be employed for obtaining samples from this
joint posterior distribution. We used a
Metropolis-Hastings algorithm to yield the required realizations and
adopted a sampling scheme where the parameters are
updated as a single block. At each iteration we generated new
parameter values
from a multivariate normal distribution centred around the maximum
likelihood estimates with a variance-covariance proposal matrix given by the inverse
Hessian evaluated at the posterior mode. This variance-covariance
matrix was then tuned so that the acceptance rates were between 0.3
and 0.6.

\subsection{Bayesian prediction in GARMA models}

The Bayesian model is defined by equation (\ref{jjj1}) where 
information from the observed data is combined through the likelihood
function with prior information.
In practice however, the interest is often
in future values of the series which are probabilistically represented
by the $h$-steps ahead predicitive distribution. Denoting by 
$\bftheta=(\beta_0,\Phi,\Theta,u,\lambda)$ the set of all parameters
in the model, this predictive density is given by,
\begin{equation*}
f(y_{t+h}|F_t) = \int f(y_{t+h}|\bftheta,F_t)\pi(\bftheta|F_t)d\bftheta.
\end{equation*}

\noindent and a point prediction is given by its expectation,
\begin{eqnarray*}
E(y_{t+h}|F_t) &=& \int_{y_{t+h} \in Y_{t+h}}y_{t+h} f(y_{t+h}|F_t)dy_{t+h}\\
&=&
\int_{y_{t+h} \in Y_{t+h}}y_{t+h}\left[\int_{\theta\in\Theta}f_{\Theta}(y_{t+h}|F_t)\pi(\bftheta|F_t)d\theta\right]
dy_{t+h}\nonumber\\
&=&
\int_{\theta\in\Theta} 
\left[\int_{y_{t+h} \in Y_{t+h}}y_{t+h} f_{\Theta}(y_{t+h}|F_t) dy_{t+h}\right]\pi(\bftheta|F_t) d\bftheta\\
&=&
\int_{\theta \in \Theta} \left[ E(y_{t+h}|F_t,\bftheta)\right] \pi(\bftheta|F_t) d\bftheta = \mu_{t+h}(\bftheta).
\end{eqnarray*}

\noindent Then, given a sample $\bftheta^{(1)},\dots,\bftheta^{(Q)}$ from the
posterior distribution of $\bftheta$, this expectation is approximated as,
\begin{equation*}
\hat{E}(y_{t+h}|F_t) = \hat{y}_{t+h}= \frac{1}{Q}\sum_{l=1}^Q\mu_{t+h}(\theta^{(l)}),
\end{equation*}
where $\mu_{t+h}^{(l)}$ is obtained at each MCMC iteration using the
link function, i.e. 
\begin{equation}\label{pred}
g(\mu_{t+h}^{(l)}) = 
\beta_0^{(l)} + \sum_{i=1}^p\phi_{i}^{(l)}g(y_{t+h-i}) + \sum_{j=1}^q\theta_{j}^{(l)}[g(y_{t+h-j})-g(\mu_{t+h-j})]
\end{equation}

\noindent We also note that, since
\begin{displaymath}
E(y_{t+h-j}|F_t) = \left\{ \begin{array}{ll}
 y_{t+h-j}, h\leq j \\
 \hat{y}_{t+h-j},  h>j
\end{array} \right.
\end{displaymath}

\noindent then it follows that in (\ref{pred}),
\begin{displaymath}
\mu_{t+h-j} = \left\{ \begin{array}{ll}
 \hat{\mu}_{t+h-j}, h\leq j \\
 \hat{y}_{t+h-j},  h>j
\end{array} \right.
\end{displaymath}

However, these represent predictions for the transformed series, while
in practice one would be interested in forecasting the original
series. Since $E[y_t|F_t] =
E[\lambda(y^{(\lambda)}_t+1)^\frac{1}{\lambda}|F_t]$, we can apply the inverse
transformation on the estimated mean, thus obtaining the original
predictions as 
$\mu_{t+h}^{(l)}=[\lambda\hat{\mu}_{t+h}^{(l)} +
  1]^{\frac{1}{\lambda}}$ for each iteration. Note that these
predictions are obtained without any assumption or theoretical
expansion which provides a crucial gain of the Bayesian approach.
Prediction intervals can also be obtained using 
the MCMC sample to calculate $\mu^{(l)}_{t+h} =
g^{-1}\left(\eta_{t+h}(\gamma^{(l)})\right)$,
$l=1,\ldots,M$ and computing the associated lower and upper quantiles.

\section{Simulation Studies}\label{simulations}

In this section we present simulation studies to examine the
performance of Bayesian estimation and Bayesian model selection in
TGARMA models. 
The performance of the Bayesian estimation was evaluated using three
metrics: the corrected bias (CB), the corrected error (CE) and the
mean acceptance rates in the MCMC algorithm called acceptance
probabilities (AP). These metrics are defined as,
\begin{eqnarray*}
CB &=& \frac{1}{m} \sum_{i=1}^m
\left|\frac{\theta-\hat{\theta}^{(i)}}{\theta}\right|,\label{bias}\\
CE^2 &=&
\frac{1}{\tau^2}\frac{1}{m}\sum_{i=1}^m(\hat{\theta}^{(i)}-\theta)^2\label{mse}\\
AP &=& \frac{1}{m} \sum_{i=1}^m \hat{r}^{(i)}\label{ap},
\end{eqnarray*}
where $\hat{\theta}^{(i)}$  and $\hat{r}^{(i)}$ are the estimate of
parameter $\theta$ and the computed acceptance rate, respectively, for
the $i$-th replication, $i=1,\ldots,m$. In this paper we take the
posterior means of $\theta$ as point estimates. Furthermore, the variance
term $\tau^2$ that appears in the definition of CE is the sample
variance of $\hat{\theta}^{(1)},\dots,\hat{\theta}^{(m)}$.

To simulate the artificial data we specified parameter values that would generate
moderate values for the time series. Also, to investigate the
robustness of the Bayesian estimation to the transformation parameter
we conducted this study with different values of $\lambda$. 
In all cases, the experiment was replicated $m=1000$ times for each model. 
For each
dataset, we used the prior distributions as described in Section \ref{sec:bayes}
with mean zero and variance 200. We then drew samples from the
posterior distribution 
discarding the first 1000 draws as burn-in and keeping every 3rd
sampled value, resulting in a final sample of 5000 values. 
we used the diagnostic proposed by \cite{geweke} to
assess convergence of the chains. This is based on a test for
equality of the means of the first and last parts of the chain (by
default the first 10$\%$ and the last 50$\ $). If the samples are
drawn from the stationary distribution, the two means are equal and
the statistic has an asymptotically standard normal distribution. The
calculated values of Geweke statistics were all 
between -2 and 2, which is an indication of convergence of the Markov
chains. All the
computations were implemented using the open-source statistical
software language and environment {\tt R} (\cite{r10}). 

Tables \ref{tab1} and \ref{tab3} show the results obtained 
for TGARMA(1,1) and TGARMA(2,2) models following a Gamma
distribution with $\lambda\in\{0.3,0.5,0.7,0.9\}$. These tables show
acceptable values for CB, CE and AP, which should be near 0, 1 and
between 0.30 and 0.80, respectively. It is worth noting that
similar results were obtained with TGARMA(1,2) and TGARMA(2,1) models
and that we also conducted the same study using the inverse
Gaussian distribution and again similar results were
obtained.

In terms of model selection performance, Table \ref{tab2} presents the
proportions of correct models chosen using Bayesian criteria with
Gamma GARMA($p,q$) models and varying
$\lambda\in\{0.3,0.5,0.7,0.9\}$. The criteria used are the DIC
(deviance information criterion), EBIC (expected BIC) and CPO 
(conditional predictive ordinate).
We note that the CPO criterion is not
selecting particularly well the TGARMA(1,1) for any value of
$\lambda$. Likewise for the EBIC with the TGARMA(2,2) models. The DIC
however has acceptable proportions of correct model for all
combinations of TGARMA and $\lambda$.

\section{Real data analysis}\label{real}

In this section the methodology described in the previous sections is
applied to a real time series data.
The demography of Sweden is monitored by Statistics Sweden (SCB). As
of 31 December 2013, Sweden's population was estimated to be 9.64
million people, making it the 90th most populous country in the
world. The three largest cities are Stockholm, Gothenburg and
Malm\"o. Approximately 85$\%$ of the country's population resides in
urban areas. 

\begin{figure}[h!]\centering
\includegraphics{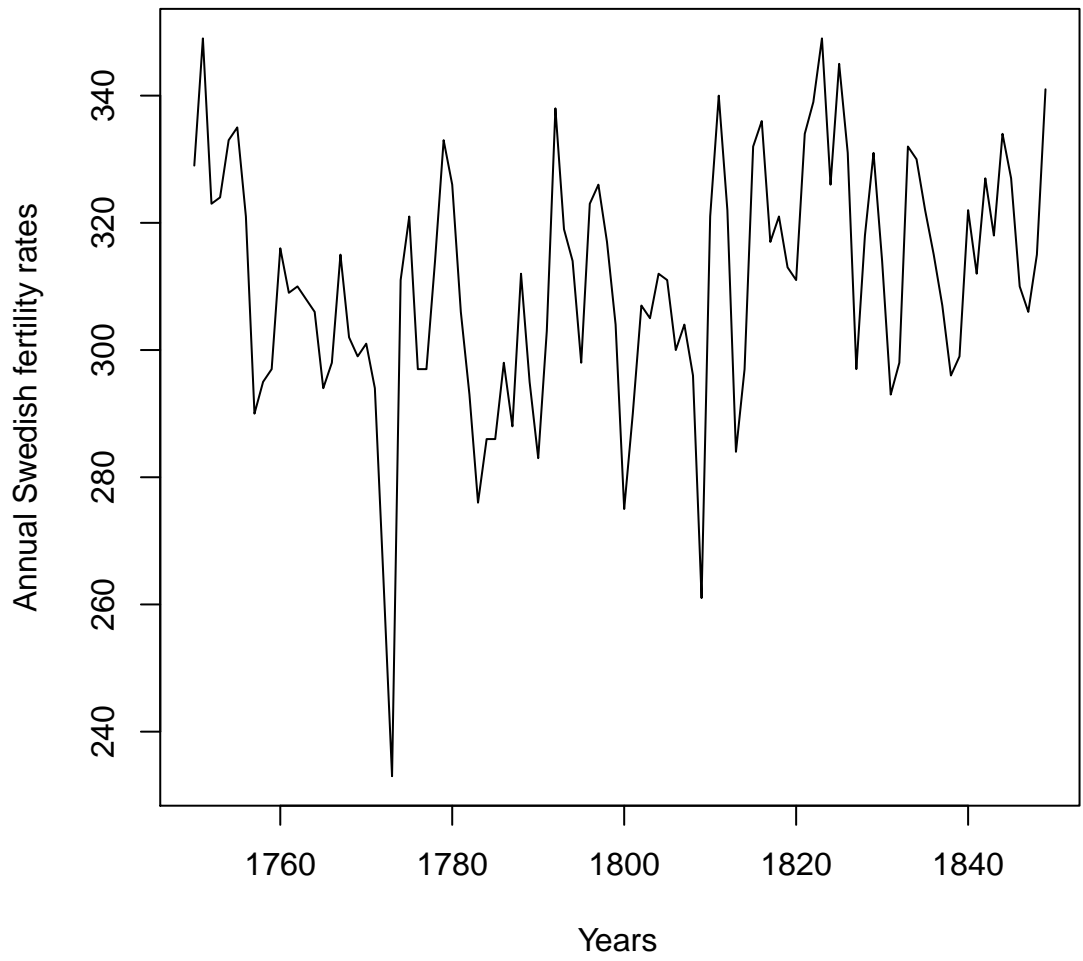}
\caption{Annual Swedish fertility rates from 1750 to 1849.}
\label{fig1}
\end{figure}

\begin{figure}[h!]\centering
\includegraphics{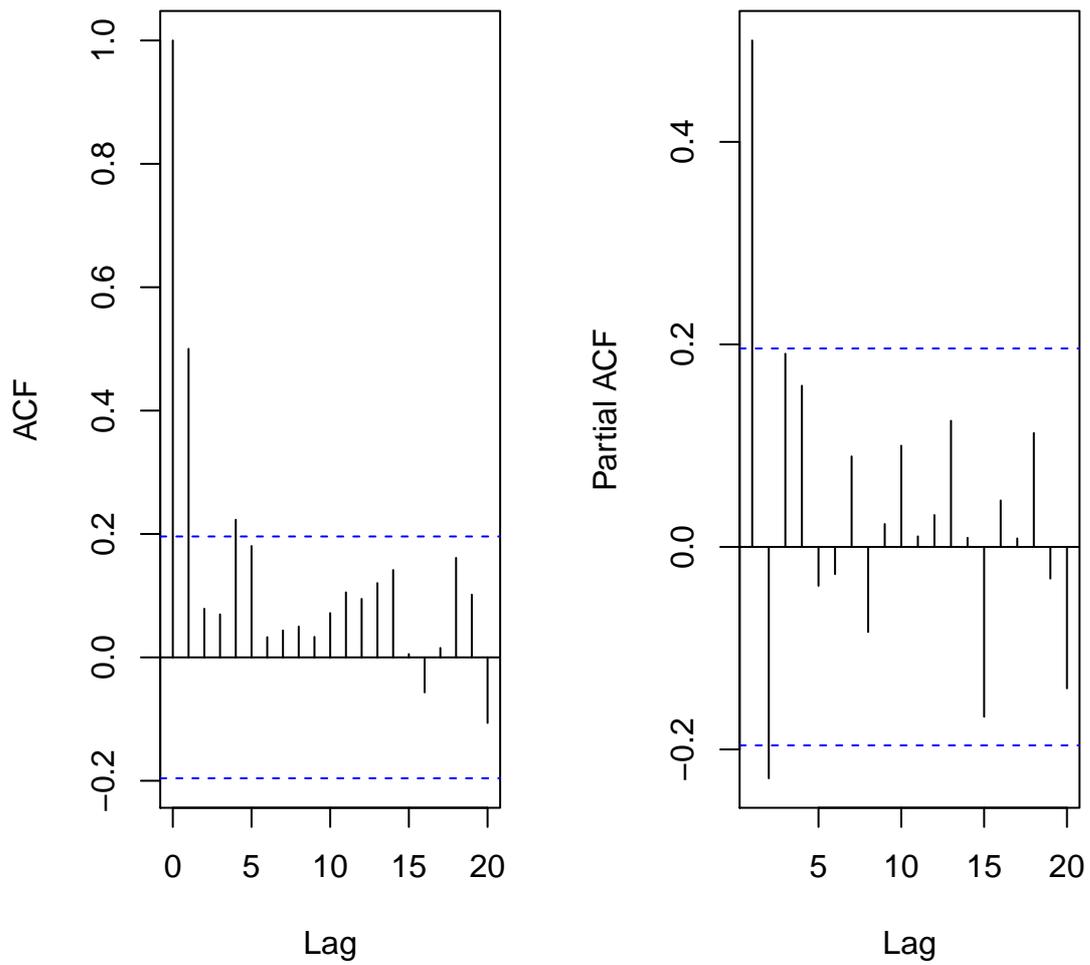}
\caption{Autocorrelations and partial autocorrelations of the Annual Swedish fertility rates.}
\label{fig2}
\end{figure}

The real data set to be analysed refers to the Annual Swedish fertility rates
(1000's) from 1750 to 1849. These data are depicted in Figure
\ref{fig1} and were obtained from the website
{\tt https://datamarket.com/data/set/22s2}.
Also, Figure \ref{fig2} presents the sample autocorrelations and
partial autocorrelations for the Annual Swedish fertility rates
respectively.

In the first step, Bayesian selection criteria were used to compare between gamma
and inverse Gaussian models and also to select the model order. The
same criteria DIC, EBIC and CPO were employed and Table \ref{tab4}
presents the results. From this table we can see that the TGARMA(1,0)
model with gamma distribution is preferred in terms of all the
criteria although the TGARMA(1,1) with the same distribution is a
close competitor.
Table \ref{tab5} presents the Bayesian
estimates of the selected model with posterior means and standard deviations (in
brackets), the 95\% HPD intervals and acceptance rates from the
Metropolis algorithm. 

\begin{figure}[h!]\centering
\includegraphics[width=0.6\linewidth,angle=270]{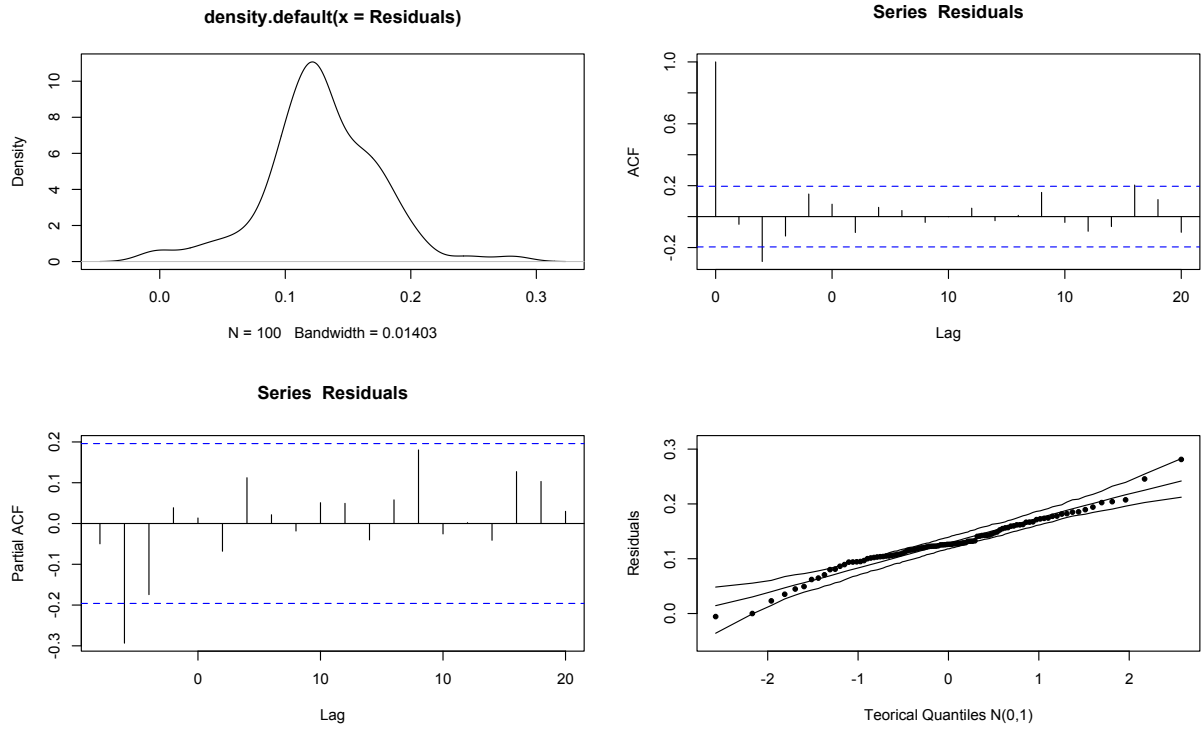}
\caption{Autocorrelation function and partial autocorrelation function
  of the residuals of the Annual Swedish fertility rates series.} 
\label{fig4}
\end{figure}

Finally, a residual analysis was carried out to assess the adequability
of the chosen model.
Quantile residuals are based on the idea of inverting the estimated
distribution function for each observation to obtain exactly standard
normal residuals. This is accomplished by defining the residuals as
$r_t=\Phi^{-1}(\textbf{F}_{y_t}(y_t|F_{t-1}))$ where
$\textbf{F}_{y_t}$ represents the cumulative distribution function for
the associated density funcion. Figure \ref{fig4} confirms residuals following
Gaussian distribution and non-correlated.

\begin{figure}[h!]\centering
\includegraphics[width=0.5\linewidth,angle=270]{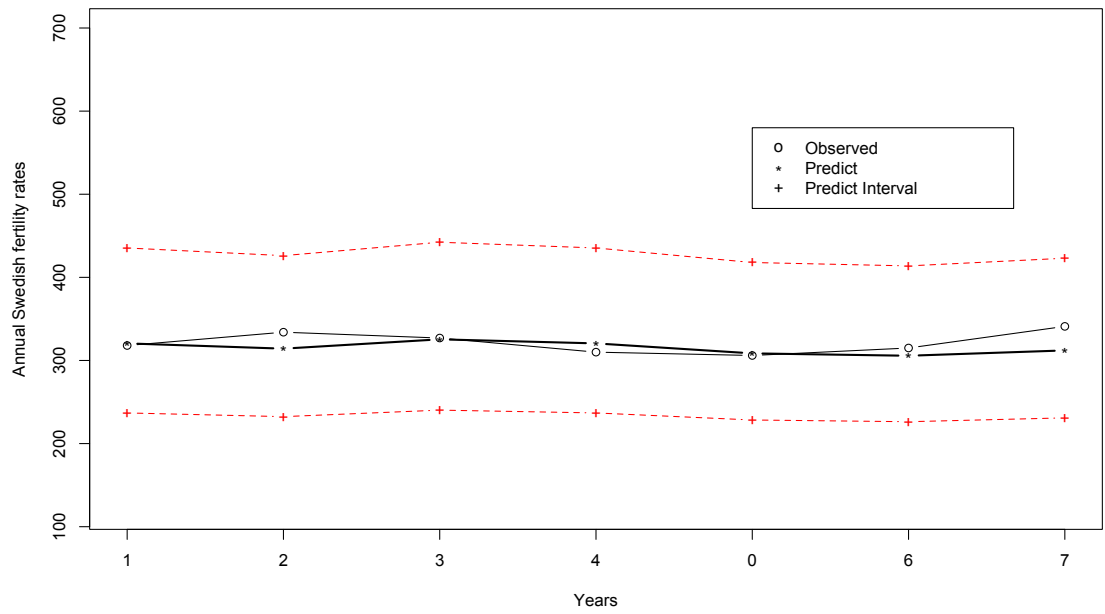}
\caption{Predictions with TGARMA(1,0) Gamma model for the Annual Swedish fertility rates series.}
\label{fig5}
\end{figure}

The prediction were made by the median. Only the first term of Taylor
expansion was used. Using the estimate, predictions of 6 steps ahead
of the original series can be made. The 6 last values of the series
were removed and fitted the model without them. Figure \ref{fig5}
presents predictions one step ahead for 6 years values, thus the
predicted value be compared with the true value. The MAPE was
calculated to assess the quality of predictions, the value was
$03.70\%$ which indicated good predictions.

\section{Discussion}\label{conclusions}

In this paper we discussed a Bayesian approach for estimation,
comparison and prediction of TGARMA time series models. We analyzed
two different continuous models: gamma and inverse Gaussian. We
implemented MCMC algorithms to carry out the simulation study and the
methodology was also applied on a real time series dataset.

Properties of the Bayesian estimation and the performance of Bayesian
selection criteria were also assessed with our simulation study. The
analysis with real data also provided good estimates and predictions
via parsimonious models. Our results suggest that, as indicated in the
original TGARMA paper, this class of models has potential uses for
modeling non-additivity, non-normality and heteroscedasticity
continuous time series.

\section*{Acknowledgements}

Breno Andrade gratefully acknowledges the financial support from Brazilian research agency CAPES.
Ricardo Ehlers received support from S\~ao Paulo Research Foundation
(FAPESP) - Brazil, under grant number 2015/00627-9. The authors
gratefully acknowledge the comments and constructive suggestions by an anonymous referee.


\clearpage

\begin{table}[ht!]
\caption{Corrected bias (CB), corrected error (CE) and the mean
  acceptance rates (AP) for the TGARMA(1,1) model with gamma
  distribution and Box-Cox power transformation.} 
\label{tab1}
\begin{center}
\begin{tabular}{c|c|c|c|c|c|c}
\hline
Parameter   & True value &  Mean   &  Variance  & CB       & CE        & AP          \\ 
 \hline 
 $\lambda $ & 0.30       & 0.3025  &  0.0039    & 0.1504   & 0.9982    & 0.5939    \\ 
 $\nu $     & 0.50       & 0.5032  &  0.0011    & 0.0528   & 1.0023    & 0.7414    \\ 
 $\alpha_0$ & 0.70       & 0.6970  &  0.0277    & 0.1793   & 0.9976    & 0.6302     \\ 
 $\alpha_1$ & 0.50       & 0.4970  &  0.0019    & 0.0718   & 0.9996    & 0.5553    \\     
 $\phi_1$   & 0.30       & 0.3008  &  0.0016    & 0.1036   & 0.9976    & 0.5863      \\ 
 \hline
 $\lambda $ &   0.50     & 0.5052  &  0.0068    &  0.1266   &   1.0015      &  0.6771    \\ 
 $\nu $     &   0.50     & 0.5030  &  0.0010    &  0.0511     &   1.0040     &  0.7455    \\ 
 $\alpha_0$ &   0.70     & 0.7027  &  0.0281    &  0.1882       & 0.9996         &  0.6370    \\ 
 $\alpha_1$ &   0.50     & 0.4997  &  0.0018    &  0.0689     &   0.9995      &  0.5574    \\     
 $\phi_1$   &   0.30     & 0.3008  &  0.0016    &  0.1069      &   1.0015      & 0.5854      \\ 
\hline
 $\lambda $ &   0.70     & 0.7095  &  0.0110      &  0.1208   &   1.0006      &  0.7276    \\ 
 $\nu $     &   0.50     & 0.5076  &  0.0013      &  0.0598   &   1.0182      &  0.7450    \\ 
 $\alpha_0$ &   0.70     & 0.7011  &  0.0316       &  0.2019   &   0.9964       &  0.6384     \\ 
 $\alpha_1$ &   0.50     & 0.4978  &  0.0019      &  0.0722  &   0.9976       &  0.5611    \\     
 $\phi_1$   &   0.30     & 0.3008  &  0.0015      &  0.1096   &   0.9966      & 0.5873      \\ 
\hline
$\lambda $ & 0.90       & 0.8783  &  0.0107    &  0.0969  &   1.0203     & 0.7634    \\ 
$\nu $     & 0.50       & 0.5114  &  0.0010    &  0.0550  &   1.0581     & 0.7401    \\ 
$\alpha_0$ & 0.70       & 0.6581  &  0.0217    &  0.1778  &   1.0382     & 0.6270     \\ 
$\alpha_1$ & 0.50       & 0.4925  &  0.0018    &  0.0685  &   1.0140     & 0.5553    \\     
$\phi_1$   & 0.30       & 0.3020  &  0.0019    &  0.1147  &   0.9991     & 0.5831      \\ 
\hline
\end{tabular}
\end{center}
\end{table}

\clearpage

\begin{table}[ht!]
\caption{Proportions of correct model choice using Bayesian criteria with Gamma GARMA($p,q$) models.}
\label{tab2}
\begin{center}
\renewcommand{\arraystretch}{1}

\begin{tabular}{c|c|c||c|c}
\hline
\multicolumn{1}{c|}{\textbf{ }}&\multicolumn{2}{c||}{$\lambda=0.3$}&\multicolumn{2}{c}{$\lambda=0.5$}\\
  \hline
  Model      & GARMA(1,1)   &     GARMA(2,2)   & GARMA(1,1)   &   GARMA(2,2)   \\ 
  \hline
   EBIC     & 0.9820        &     0.4640          & 0.9920           &     0.4220         \\ 
   DIC      & 0.7900        &     0.7660          & 0.7940           &     0.7760         \\ 
   CPO      & 0.4260        &     0.7860          & 0.4300           &     0.8040          \\ 
  \hline 
\end{tabular}

\begin{tabular}{c|c|c||c|c}
\multicolumn{1}{c|}{\textbf{ }}&\multicolumn{2}{c||}{$\lambda=0.7$}  & \multicolumn{2}{c}{$\lambda=0.9$}\\
  \hline
  Model     & GARMA(1,1)   &      GARMA(2,2)   & GARMA(1,1)   &   GARMA(2,2)   \\ 
  \hline  
   EBIC     & 0.9880       &     0.4900        & 0.9890       &     0.4540         \\ 
   DIC      & 0.8080       &     0.7860        & 0.7800       &     0.7510          \\ 
   CPO      & 0.4800       &     0.7800        & 0.4860       &     0.7950          \\ 
  \hline 
\end{tabular}
\end{center}
\end{table}

\clearpage

\begin{table}[ht!]
\caption{Corrected bias (CB), corrected error (CE) and the mean
  acceptance rates (AP) for the TGARMA(2,2) model with gamma
  distribution and Box-Cox power transformation.} 
\label{tab3}
\begin{center}
\renewcommand{\arraystretch}{1}
\begin{tabular}{c|c|c|c|c|c|c}
\hline
Parameter   & True value &    Mean     &  Variance    &  CB           & CE  &  AP          \\ 
\hline 
$\lambda $  &   0.30    &  0.3013  &  0.0010     &  0.0762    &  0.9971   &  0.3535  \\ 
$\nu $      &   0.50    &  0.5060  &  0.0008     &  0.0474    &  1.0170   &  0.7407    \\ 
$\alpha_0$  &   0.50    &  0.5332  &  0.0398     &  0.3074    &  1.0100   &  0.4085     \\ 
$\alpha_1$  &   0.30    &  0.2790  &  0.0178     &  0.3406    &  1.0084   &  0.2771     \\     
$\alpha_2$  &  -0.20    & -0.2064  &  0.0020     &  0.1810    &  1.0062   &  0.5685     \\   
$\phi_1$    &   0.40    &  0.4184  &  0.0182     &  0.2566    &  1.0055   & 0.2478     \\ 
$\phi_2$    &  -0.30    & -0.2788  &  0.0140     &  0.2956    &  1.0121   & 0.2421     \\ 
\hline
$\lambda $       &   0.50       &  0.5031  &  0.0020      &  0.0654      &   0.9989    &  0.4252  \\ 
$\nu $          &   0.50      &  0.5043   &  0.0009      &  0.0498      &   1.0063     &  0.7521    \\ 
$\alpha_0$   &   0.50      & 0.5289    &  0.0380      & 0.3019       & 1.0076        &  0.4871     \\ 
$\alpha_1$   &   0.30      & 0.2841    &  0.0169      &  0.3319       &   1.0040    &  0.3257     \\     
$\alpha_2$   &   -0.20      & -0.2055    &  0.0020      &  0.1814       &   1.0042     &  0.5764     \\   
$\phi_1$      &   0.40      & 0.4139     &  0.0171    &  0.2461       &   1.0023      & 0.3301     \\ 
$\phi_2$    &   -0.30     & -0.2838    &  0.0133     &  0.2903       &   1.0064     & 0.3457      \\ 
\hline
$\lambda $       &   0.70       &   0.7015  &  0.0036      &  0.0660      &   0.9961    &  0.5057  \\ 
$\nu $          &   0.50      &  0.5064   &  0.0009      &  0.0501      &   1.0174     &  0.7378    \\ 
$\alpha_0$   &   0.50      & 0.5321    &  0.0406      & 0.3102       & 1.0085        &  0.4414     \\ 
$\alpha_1$   &   0.30      & 0.2822    &  0.0175      &  0.3347       &   1.0048     &  0.3271     \\     
$\alpha_2$   &   -0.20      & -0.2076    &  0.0018      &  0.1732       &   1.0118     &  0.6342     \\   
$\phi_1$      &   0.40      & 0.4153     &  0.0173    &  0.2483       &   1.0026      & 0.2621     \\ 
$\phi_2$    &   -0.30     &-0.2813    &  0.0133     &  0.2905       &   1.0089      & 0.2928      \\ 
\hline
$\lambda $      &    0.90    &  0.9025  &  0.0055    &  0.0642      &   0.9948    &  0.5932  \\ 
$\nu $       &    0.50    &  0.5068  &  0.0009    &  0.0520      &   1.0177     &  0.7285    \\ 
$\alpha_0$   &    0.50    &  0.5347  &  0.0442    &  0.3131      & 1.0079        &  0.5121     \\ 
$\alpha_1$   &    0.30    &  0.2793  &  0.0156    &  0.3093       &   1.0075     &  0.2318     \\     
$\alpha_2$   &   -0.20    & -0.2069  &  0.0016    &  0.1654       &   1.0087     &  0.5478     \\   
$\phi_1$     &    0.40    &  0.4191  &  0.0157    &  0.2304      &   1.0060      & 0.2114     \\ 
$\phi_2$     &   -0.30    & -0.2807  &  0.0126    &  0.2741       &   1.0089      & 0.2407      \\ 
\hline
\end{tabular}
\end{center}
\end{table}

\clearpage

\setlength{\tabcolsep}{1mm}

\begin{table}[ht!]
\caption{Computed values of information criteria for the Annual Swedish fertility rates.}
\label{tab4}
\begin{center}
\renewcommand{\arraystretch}{1}
\begin{tabular}{l|c|c|c|c|c}
\hline
Gamma  &  TGARMA(1,0)  & TGARMA(1,1)  &  TGARMA(1,2)    &  TGARMA(2,1)   &  TGARMA(2,2)    \\ 
\hline  
EBIC   &  1506.72      &  1507.51      &    1579.31         &  1544.57            &   1606.73              \\ 
DIC    &  1496.66      &  1497.32      &    1570.84           &   1530.09            &    1596.48        \\ 
CPO    &  -299.91      &  -300.76      &    -409.56           &   -306.09            &    -435.69    \\ 
  \hline
Inv. Gaussian  & TGARMA(1,0)    & TGARMA(1,1)   &  TGARMA(1,2)  &  TGARMA(2,1)  &  TGARMA(2,2)   \\ 
\hline  
EBIC   &  1664.14       &  1664.89      &      1683.51         &  1681.97            &   1686.55           \\ 
DIC    &  1651.29       &  1652.56      &      1667.54           &  1665.44            &   1665.82  \\ 
CPO    &  -336.96       &  -337.35      &      -381.22           &   -374.79           &    -392.01 \\
\hline 
\hline
\end{tabular}
\end{center}
\end{table}

\clearpage

\begin{table}[ht!]
\caption{Estimates of Annual Swedish fertility rates series with TGARMA(1,0) Gamma.}
\label{tab5}
\begin{center}
\renewcommand{\arraystretch}{1}
\begin{tabular}{ccccc}
\hline
Parameter  & Mean (SD) &  lower HPD limit & Upper HDP limit  & AP  \\ 
   \hline  
$\beta_0$  & 0.7888 (0.0537)   &     0.6497      &   0.9269 & 0.4928 \\ 
$\phi_1$   & 0.7157 (0.0224)   &     0.6657      &   0.7650 & 0.4961 \\ 
$\nu$      & 3.4782 (0.3516)   &     2.8195      &   4.1212 & 0.6701 \\   
$\lambda$  & 0.3145 (0.0326)   &     0.2513      &   0.3754 & 0.7382 \\   
\hline
\end{tabular}
\end{center}
\end{table}

\clearpage

\clearpage

\clearpage

\end{document}